# Mid-infrared-to-ultraviolet supercontinuum generation in low-loss tantalum pentoxide nanophotonic waveguides

Minghui Li,[1,5,11] Qiankun Li,[2,11] Xhizhi Zheng,[3,4,11] Xueying Sun,[2] Renhong Gao,[3] Fan Jiang,[6] Hairun Guo,[2,8,*] Jintian Lin, [1,5,†] and Ya Cheng[3,4,7,8,9,10]

*[1]State Key Laboratory of Ultra-intense Laser Science and Technology, Shanghai Institute of Optics and Fine Mechanics, Chinese Academy of Sciences, Shanghai 201800, China*

*[2]Key Laboratory of Specialty Fiber Optics and Optical Access Networks, Shanghai University, Shanghai 200444, China*

*[3]The Extreme Optoelectromechanics Laboratory (XXL), School of Physics, East China Normal University, Shanghai 200241, China*

*[4]State Key Laboratory of Precision Spectroscopy, East China Normal University, Shanghai 200062, China*

*[5]Center of Materials Science and Optoelectronics Engineering, University of Chinese Academy of Sciences, Beijing 100049, China*

*[6]University College London, London WC1E 6BT, United Kingdom*

*[7]Shanghai Research Center for Quantum Sciences, Shanghai 201315, China*

*[8]Hefei National Laboratory, Hefei 230088, China*

*[9]Collaborative Innovation Center of Extreme Optics, Shanxi University, Taiyuan 030006, China*

*[10]Collaborative Innovation Center of Light Manipulations and Applications, Shandong Normal University, Jinan 250358, China*

*[11]Minghui Li, Qiankun Li, and Xinzhi Zheng contributed equally to this work.*

**E-mail: hairun.guo@shu.edu.cn*

*[†]E-mail: jintianlin@siom.ac.cn*

# Abstract:

Optical frequency combs (OFCs) on photonic integrated platforms are revolutionizing precision metrology, bio-imaging, atomic and molecular sensing, and ultrafast photonics, yet most remain confined to the near-infrared. This restriction prevents access to the ultraviolet, visible, and mid-infrared bands critical for various quantum, atomic, and molecular systems. Here, we overcome this challenge by exploring tantalum pentoxide as a platform for ultra-broadband OFCs, leveraging its broad transparency window (300−8000 nm), a high nonlinear refractive index three times larger than that of silicon nitride, and a wide bandgap that suppresses two-photon absorption at short wavelengths. Critically, by using a photolithography assisted chemo-mechanical etching process, we achieve waveguides simultaneously featuring appropriate dispersion and record-low propagation losses of 0.066 dB/cm at telecom wavelength, significantly facilitating the supercontinuum spectral extension into the ultraviolet and the mid-infrared. Pumping these anomalous-dispersion waveguides with femtosecond pulses at 1550 nm yields a gap-free, 3.2-octave supercontinuum spanning from 350 to 3200 nm via a soliton-based dynamics at only 54 pJ pulse energy, representing the broadest comb spectrum on this platform. By further tailoring the waveguide dispersion to a flattened normal profile, a relatively flat spectrum with -30 dB bandwidth of 1182 nm is obtained. Moreover, excellent comb coherence is validated via heterodyne detection, and efficient soliton-effect pulse compression from 126.7 fs to 19.2 fs is achieved. This work establishes a versatile platform for chip-scale OFCs that seamlessly bridge the ultraviolet, visible, near-infrared, and mid-infrared. This breakthrough will impact quantum information processing, optical clock applications, and precision molecular spectroscopy in this important spectral window.

# Introduction

Integrated nonlinear photonic devices have revolutionized frequency conversion for generating new optical frequencies, by combining low power consumption, a small device footprint, and unprecedented nonlinear conversion efficiencies[1-3]. Among these, octave-spanning optical frequency combs (OFCs) on chip-scale photonic platforms are game-changing for precise spectroscopic metrology, high-capability communications, high-speed photonic computation, miniaturized optical clocks, and integrated quantum technologies[4-15]. To date, chip-scale OFCs are primarily focused on the near-infrared (NIR), with only initial extension into the visible and mid-infrared (mid-IR)[8,16-18]. Achieving a single integrated OFC that simultaneously spans the ultraviolet (UV), visible, NIR, and mid-IR (>2700 nm) remains a critical challenge. This limitation directly impedes multiple impactful research fields such as optical atomic clocks, quantum computing and sensing, and atomic and molecular spectroscopy, which highly require precision absolute frequency measurement in the UV and visible spectral bands for multiple trapping, cooling, and state preparation transitions of atoms and ions. It also constrains NIR-based high-bandwidth data transmission, the light detection and ranging system, high-speed optical computation, and integrated quantum technology and their practical applications[18-20]. Finally, in the fields of mid-IR spectroscopy, most molecules are responsible for absorption at specific wavelength in the mid-IR >2700 nm spectrum[4], enabling the detection and quantification of small traces of substances. The fundamental bottleneck is the lack of a photonic platform that simultaneously provides appropriate engineered dispersion, low propagation losses, a transparency window spanning from the UV to mid-IR, a moderate refractive index, a large bandgap to suppress two-photon absorption at short wavelengths, and a substantial third-order nonlinearity for accessing Kerr nonlinear processes.

Tantalum pentoxide ($Ta_2O_5$) is an ideal candidate:[21-26] it offers a low thermo-optic coefficient ($2.3\times10^{-6}$/K, lower than those of $Si_3N_4$ of $2.45\times10^{-5}$/K and $SiO_2$ of $8.74\times10^{-6}$/K), a relatively high refractive index ($n\approx2.07$), an ultra-broad transparency window from 0.33 to 8 μm (surpassing $Si_3N_4$'s 0.4–3 μm), a large bandgap of 4.3 eV, a negligible two-photon absorption coefficient in the visible, and a strong nonlinear refractive index of $7.8\times10^{-19}$ $m^2$/W which is three times higher than that of $Si_3N_4$. Moreover, $Ta_2O_5$ is compatible with room-temperature deposition and yields low-residual-stress thick films optimized for dispersion management, enabling crack-free photonic layer without the high-temperature processing required for $Si_3N_4$. These attributes make it exceptionally promising for thermo-stable, efficient chip-scale OFCs from the UV to the mid-IR. However, the extreme hardness and chemical inertness of $Ta_2O_5$ have long hindered the fabrication of high-confinement $Ta_2O_5$ waveguides with smooth sidewalls and anomalous dispersion. Typical propagation loss of 0.3 dB/cm at telecom wavelengths have precluded the formation of broad-bandwidth OFCs. A recent advance demonstrated a normal-dispersion $Ta_2O_5$ waveguide cladded with $SiO_2$, achieving a record-low propagation loss of 0.08 dB/cm[23]. However, when the waveguide top width was reduced to approximately 1.5 μm for anomalous dispersion, the propagation loss collapsed to approximately 0.3 dB/cm. Furthermore, the $SiO_2$ cladding introduces strong O-H absorption beyond 2500 nm[23], blocking mid-IR. Moreover, the high sensitivity of scattering loss to sidewall roughness at short wavelengths limited the state-of-the-art propagation loss at 780 nm to 0.86 dB/cm[26]. Consequently, chip-scale $Ta_2O_5$ OFCs have remained confined to the 750–2400 nm range[23], and the spectral extension to the UV and mid-IR is still challenging.

Here, we overcome these limitations and demonstrate ultra-broadband OFCs spanning continuously from the UV to the mid-IR in integrated $Ta_2O_5$ rib waveguides. The key innovation lies in a photolithography assisted chemo-mechanical etching process that yields an average sidewall roughness of merely 0.34 nm, drastically suppressing scattering loss across the entire spectrum while simultaneously enabling precision dispersion engineering–all without high-temperature (>500°C) annealing or an absorbing $SiO_2$ cladding. Under fundamental transverse-electric ($TE_0$) excitation, soliton fission, dispersive-wave formation, and third-harmonic generation (THG) in the anomalous-dispersion regime generate a gap-free, 3.2-octave supercontinuum from 350 nm to 3200 nm at a pump pulse energy of only 54 pJ. By further tailoring the waveguide dispersion into a flattened normal profile, we obtain a relatively flat spectrum with a -30 dB bandwidth of 1182.331 nm, showcasing the versatility of this approach for tailored spectral shaping. The excellent coherence of the generated comb is validated via heterodyne interference, pulse compression of the supercontinuum is demonstrated, and the mode-phase-matching conditions for THG are analyzed. This integrated photonic platform, uniquely combining a nanostructuring technique simultaneously providing low propagation-loss and precision dispersion management, wafer-scale crack-free $Ta_2O_5$ films, and a high laser damage threshold, establishes a route to multi-octave frequency combs that fully bridge the ultraviolet, visible, near-infrared, and mid-infrared—opening transformative opportunities for quantum information processing, optical clocks, and precision molecular spectroscopy across this vast spectral window.

## Results

### Fabrication of dispersion-engineered $Ta_2O_5$ nanophotonic waveguides with low propagation losses

The dispersion-engineered $Ta_2O_5$ rib waveguides were fabricated on a 900-nm-thick $Ta_2O_5$ film deposited magnetron sputtering on a 2-μm-thick thermal oxide layer substrate, which was supported by a 4-inch, 500-μm-thick silicon wafer. The low-loss rib waveguides were then patterned using femtosecond laser photolithography assisted chemo-mechanical etching (PLACE)[27]. The process consists of five key steps, including deposition of chromium (Cr) hard mask, direct writing of the waveguide pattern by femtosecond laser ablation of the Cr film, pattern transfer into the $Ta_2O_5$ film via chemo-mechanical polishing (CMP) for etching the exposed $Ta_2O_5$ region, removal of the Cr mask, and a secondary CMP step to thin and polish the $Ta_2O_5$ microstructures. The secondary CMP is critical: it substantially suppresses sidewall scattering loss while simultaneously allowing precise trimming of the waveguide thickness to engineer the dispersion profile[16].

The rib waveguide geometry was designed using a commercial finite-element method solver and is schematically shown in Fig. 1(b), where $W$, $D$, $H$, and $\theta$ denote the top width, rib heigh, total film thickness, and sidewall wedge angle, respectively. By adjusting these parameters, the group-velocity dispersion can be precisely tuned into the anomalous dispersion regime required for soliton-based supercontinuum generation. To further improve the fiber-chip coupling efficiency, the waveguide top width was adiabatically increased 8 μm at the end facets over a 200-μm-long length, as depicted in an optical micrograph of Fig. 1(a). The fabricated waveguide exhibits an ultra-smooth surface with an average roughness of only 0.46 nm[27], which drastically suppresses scattering loss–a prerequisite for extending the supercontinuum into the short-wavelength band and for achieving low-pump-power operation. A photograph of a 6.5-mm-long rib waveguide chip with taper-terminated facets is shown in Fig. 1(c), where bright supercontinuum generation is already visible under test.

To characterize the propagation loss of these ultra-smooth waveguide, a microring with a radius of 300 μm was also fabricated side-coupled with a bus waveguide using the identical PLACE process. A coupling gap was chosen to operate in the over-coupled regime, and a scanning-electron-microscope (SEM) image of the microring is shown in Fig. 1(d). By scanning a narrow-linewidth tunable laser across a fundamental transverse-electric ($TE_0$) resonance in the telecom C-band, a loaded Q factor $Q_L$ = 3.74×10$^6$ was measured (Fig. 1(e)). The corresponding intrinsic Q factor was inferred to be $Q_i$ = 5.75×10$^6$, from which the propagation loss is extracted as $\alpha$ = 0.066 dB cm$^{-1}$ (by the relation $\alpha = \frac{2\pi n_g}{Q_i\lambda}$, where $n_g$ is the group speed of 2.17, and $\lambda$ is the resonant wavelength of 1551.11 nm)[28]. Both the intrinsic Q factor and the propagation loss reported here surpass the best values previously achieved on the same platform in the telecom band[21], namely $Q_i$ = 4.5×10$^6$ and $\alpha$ = 0.09 dB/cm. Notably, at 780 nm, the loaded Q factor at 780 nm was measured to be 1.70×10$^6$ (Fig. 1(f)), yielding an intrinsic Q factor of $Q_i$ = 1.83×10$^6$ and a propagation loss of $\alpha$ = 0.43 dB cm$^{-1}$. This represents a two-fold improvement over the state of the art on the same material platform[26]. The suppressing the propagation loss to such low levels, which was achieved without high-temperature (>500°C) annealing or lossy $SiO_2$ upper cladding, not only minimizes the pump power required for supercontinuum generation, but also removes a critical barrier to extending the spectral coverage into short-wavelength regime.

## Experimental setup for supercontinuum generation

The experimental configuration for the broadband supercontinuum generation in the rib waveguides is illustrated in Fig. 2. A mode-locked pulsed laser delivering 126.7 fs pulses centered at 1550 nm with a repetition rate of 80.6 MHz, and transverse-electric (TE) polarization was used as the pump source. The pump beam passed through a half-wave plate (HWP) and a polarizing beam-splitter (PBS) for precise power management, and was then launched into the waveguide with an aspheric lens. The coupling loss between the aspheric lens and the tapered end facet was approximately 7.7 dB. For spectral analysis, the generated supercontinuum was collected from the output facet of the waveguide with a lensed fiber, and directed to one of three optical spectrum analyzers (Models: AQ6315A & AQ6375, Yokogawa Inc.; Model: OSA305, Thorlabs Inc.). The detailed procedure for the comb coherence measurement can be found in the Methods.

### Broadband supercontinuum spanning from the UV to the mid-IR

To generate a supercontinuum bridging the UV, visible, NIR, and mid-IR band, the rib waveguide was engineered with anomalous dispersion in the telecom band and designed to excite a dispersive wave generation near 940 nm. The waveguide length was 6.5 mm, with a top width of 3.05 μm, a rib height of 587 nm, a total thickness of 733 nm, and a wedge angle of 4.446°. When the on-chip average pump power reached 4.35 mW, corresponding to an average pulsed energy of 54.0 pJ, brilliant visible light was clearly observed by naked eyes along the waveguide. Figure 1(c) shows a photograph of the scattered light from soliton fission, captured by a smart-phone camera. Here, green third-harmonic light appears first, followed by yellow and orange scattering from fissioned soliton. Near the waveguide output, a clear supercontinuum with a pronounced blue component was observed, as shown in the top-view optical micrograph (left inset of Fig. 3(a)), and was projected onto a white paper screen, as demonstrated in the right inset of Fig. 3(a). The recoded spectrum (Fig. 3(a)) exhibits an ultra-broad, gap-free spectral envelope extending from 350 nm to 3200 nm, corresponding to a span of 3.195 octaves. Although the signal level drops beyond 2700 nm, the spectrum still reaches 3200 nm – a direct consequence of employing a $SiO_2$-cladding-free low-loss $Ta_2O_5$ waveguide that eliminates the strong O-H absorption typically truncating mid-IR extension. This result simultaneously pushes the short-wavelength edge into the UV and the long-wavelength edge deep into mid-IR band, representing the broadest supercontinuum ever reported on this platform.

The pump pulse energy dependence of the spectral envelope of the supercontinuum, shown in Fig. 3(b), reveals the underlying broadening dynamics. At a pump pulse energy of 15.2 pJ, self-phase modulation (SPM) broadens the spectrum from 930 nm to 2500 nm. By increasing the pump pulse energy to 21.5 pJ, the spectrum extends to a range spanning from 640 nm to 2700 nm. Here, a distinct signal emerges at 940 nm should be attributed to the excitation of a short-wavelength dispersive wave (DW), and discrete third-harmonic peaks near 570 nm begin to appear. At 26.3 pJ, the spectrum further extends to 330 nm and covers the 330–600 nm, though a spectral gap remains between 600 nm and 640 nm. At the maximum pump pulse energy of 54.0 pJ, this gap closes completely, yielding the seamless, 3.2-octave supercontinuum displayed in Fig. 3(a).

## Flattened supercontinuum via dispersion engineering

In addition to maximizing the spectral bandwidth, the waveguide geometry can be tailored to produce a spectrally flattened output. By adjusting the top width to 2.17 μm top width and the rib height to 623 nm, the integrated dispersion is shaped into a flattened normal profile. The resulting spectrum, measured at a pulse energy of 23.3 pJ, is shown in Fig. 3(c). Its –30 dB bandwidth reaches 1182.33 nm, spanning from 979.51 nm to 2161.84 nm. The insets of Figs. 3(c) and 3(d) present photographs of soliton fission scattering captured from different angles, showing blue, red, and yellow emission. This agile spectral shaping, achieved purely through dispersion engineering in a single material platform, highlights the versatility of low-loss $Ta_2O_5$ waveguides in practical applications that demand tailored spectral envelopes.

## Coherent validation and pulse compression

The coherence of the generated supercontinuum was assessed by extracting a narrow spectral band around 980 nm by band-pass filters and mixing it with an external single-frequency reference laser (black curve in Fig. 3(c)). The heterodyne beat-note spectrum, as illustrated in Fig. 3(d), contains three signals within a 100 MHz span at a resolution bandwidth (RBW) of 3.9 kHz. The strongest beat-note at 80.6 MHz ($f_{\text{repetition}}$) corresponds exactly to the repetition rate of the pump laser, while the other two singles located at 30.6 MHz and 50.0 MHz, arise from the mixing of the reference laser with the two comb lines spectrally closest to 980 nm. Crucially, the sum of these two beat frequencies of $f_{\text{beat,1}}$ and $f_{\text{beat,2}}$ (30.6 + 50.0 = 80.6 MHz) equals to the repetition rate, confirming that the supercontinuum faithfully inherits the comb structure of the pump laser. Either beat note exhibits a narrow linewidth of 3.9 kHz (inset of Fig. 3(d)), matching the linewidth of the external reference laser, and thereby verifying the excellent coherence of the supercontinuum in the NIR.

During spectral broadening, soliton self-compression occurs naturally in the waveguide. For pulses with low energy, we did not employ frequency doubling autocorrelation scheme. Alternatively, we characterized the resulting pulse compression using a FTIR spectrometer (OSA305)[29], which records the field autocorrelation trace without requiring a frequency-doubling autocorrelator. The field autocorrelation signals of the input pump pulse and the compressed output pulse are shown in Figs. 5(a) and 5(b), respectively. Fitting these spectral envelopes with Gaussian profiles, yields a pump pulse duration of 126 fs (scaling factor 0.648) and an output pulse duration of 19.2 fs, corresponding to a compression ratio of 6.56. This efficient soliton-effect compression underscores the high nonlinearity and low loss of the engineered waveguide.

**Dispersion analysis and numerical modeling**

The second-order group-velocity dispersion (GVD) and the integrated dispersion for the two characteristic waveguides, namely the one generating the 3.2-octave supercontinuum and the one producing the flattened spectrum, are plotted in Figs. 4(b) and 4(c). The anomalous-dispersion waveguide features a zero-dispersion wavelength around 940 nm, which phase-matches a DW that extends the spectrum deep into the UV and visible. While the second waveguide exhibits an ultra-flattened integrated dispersion in the normal regime, enabling the generation of a flat-envelope supercontinuum.

To quantitatively verify the spectral broadening mechanisms, we modeled the supercontinuum generation using the nonlinear Schrödinger equation (NLSE) solved by split-step Fourier method[30]:

$$\frac{\partial A(z,T)}{\partial T} + \frac{\alpha}{2} A(z,T) - \sum_{m\geq 2} \frac{i^{m+1}}{m!} \beta_m \frac{\partial^{\mathrm{m}+1} A(z,T)}{\partial t^{\mathrm{m}}} = i\gamma\left(|A|^2 A + \frac{i}{\omega_0} \frac{\partial(|A|^2 A)}{\partial T}\right). \quad (1)$$

Here, the amplitude $A(z,T)$ of the evolving pulse is a function of propagation distance $z$ and delay time defined by $T = t - z/v_g$, with $v_g$ of the group velocity at the pump wavelength. $\beta_m$ is the higher-order dispersion coefficient term in a Taylor series expansion of the propagation constant around the pump frequency $\omega_0$. In the numerical simulations, the nonlinear refractive index was set as $n_2 = 7.83\times10^{-19}$ $\mathrm{m^2/W}$ and the pulse peak powers were taken as 7 kW. Substituting these parameters in Eq. (1), yields the simulated spectra shown in Fig. 4(a), with a spectrum simulated window extending from 330 nm to 3500 nm. These simulated spectra closely match the experimentally observed ones, validating that the broad, gap-free supercontinuum in the anomalous-dispersion waveguide arises from the interplay of soliton fission, dispersive wave generation, and the SPM, whereas the flattened spectrum in the normal-dispersion waveguide is predominantly shaped by soliton fission and the SPM.

Figures 4(d) and 4(e) display the temporal and spectral evolution of the pulse along the anomalous waveguide. In the initial stage of propagation, symmetrical spectral broadening and temporal compression occur simultaneously, which are characteristic signatures of the SPM[31]. At a propagation distance of approximately 1 mm, the pulse is compressed to its shortest duration and then breaks up into several sub-pulses, making the onset of soliton fission. This clear transition from SPM-dominated broadening to soliton fission validates the nonlinear dynamics underlying the ultra-broadband supercontinuum generation in our platform.

In addition, several distinct THG peaks appear in the 400-600 nm wavelength range in Fig. 3(a). Their spectral positions are determined by phase-matching conditions between the fundamental pump mode and high-order third-harmonic modes. Two prominent THG peaks are located near 570 nm and 590 nm. The phase-matched THG processes were identified by mode analysis. Effective refractive indices of the fundamental pump mode and high-order third-harmonic modes as functions of wavelength are plotted in Fig. 4(f). Phase matching is achieved between the fundamental pump ($TE_0$) mode and the $TE_7$ and $TE_4$ modes of the third harmonic waves, corresponding precisely to the observed peaks at 570 nm and 590 nm. Other sharp peaks below 500 nm could be attributed to the phase-matched THG processes between the $TE_0$ pump mode and even higher-order third-harmonic modes.

**Discussion and Conclusion**

As a proof-of-concept demonstration, we employed a $Ta_2O_5$ rib waveguide supported on a silica substrate. The absorption of silica at longer mid-IR wavelengths currently limits further spectral extension beyond 3.2 μm. Nevertheless, this limitation is not fundamental to the waveguide platform itself. Thanks to the low residual stress inherent in thick $Ta_2O_5$ films, the same waveguide architecture can be transferred onto a sapphire substrate, which is transparent throughout the mid-IR. Adopting a sapphire substrate would remove the substrate-induced absorption bottleneck and is highly likely to push the supercontinuum much deeper into the mid-IR of 4.5 μm. In addition, the observation of distinct THG peaks within the supercontinuum offers a direct way for *f*-3*f* self-referencing, a key capability for fully stabilized frequency combs.

In summary, this work achieves the broadest supercontinuum generation ever reported in $Ta_2O_5$ nanophotonic waveguides, as summarized in Table 1. Crucially, our waveguides attain an ultra-smooth surface without any $SiO_2$ upper cladding, thereby suppressing propagation loss and strong O-H absorption that previously restricted spectral extension into both the UV and the mid-infrared. As a result, a single telecom-band femtosecond pump drives a continuous spectrum that stretches directly from the UV to the mid-IR. This unprecedented spectral coverage is enabled by the synergy of the SPM effect, dispersive wave formation, and THG within a dispersion engineered waveguide that combines a high nonlinear index, a record-low propagation loss, and a transparency window spanning from deep-UV to the mid-IR. This demonstration in which a telecom-band femtosecond laser directly drives a supercontinuum covering the UV, visible, NIR, and the mid-IR, establishes an efficient paradigm for compact, low-complexity optical frequency combs. Such a platform opens transformative opportunities for diverse photonic applications, ranging from portable optical clocks and molecular precision spectroscopy to chip-scale quantum systems.

**Table 1. Reported supercontinuum generation in $Ta_2O_5$ waveguides**

| Clad | Propagation loss | Pump (nm) | Spectrum coverage (nm) | Octave | Ref. |
|---|---|---|---|---|---|
| $SiO_2$ | 0.30 dB/cm | 1550 | 750–2400 | 1.68 | [23] |
| $SiO_2$ | ~1.0 dB/cm | 1056 | 842–1462 | 0.83 | [32] |
| $SiO_2$ | 3–4 dB/cm | 1000 | ~730–1450 | 0.99 | [33] |
| Air | ~1.0 dB/cm | 1056 | 585–1697 | 1.53 | [34] |
| Air | 0.066 dB/cm | 1550 | 350–3200 | 3.20 | This work |

## Methods

For coherence characterization, the supercontinuum output from the waveguide was first collected by an objective lens in free space, and then spectrally filtered using a band-pass filter centered at 980 nm with a 50 nm bandwidth. A narrow-linewidth (of 200 kHz) continuous-wave fiber laser operating at 980 nm serves as the reference laser. The collimated reference beam was combined with the filtered supercontinuum signal on a high-speed photodetector, and the resulting ratio-frequency microwave heterodyne beat note was recorded for coherence analysis[4].

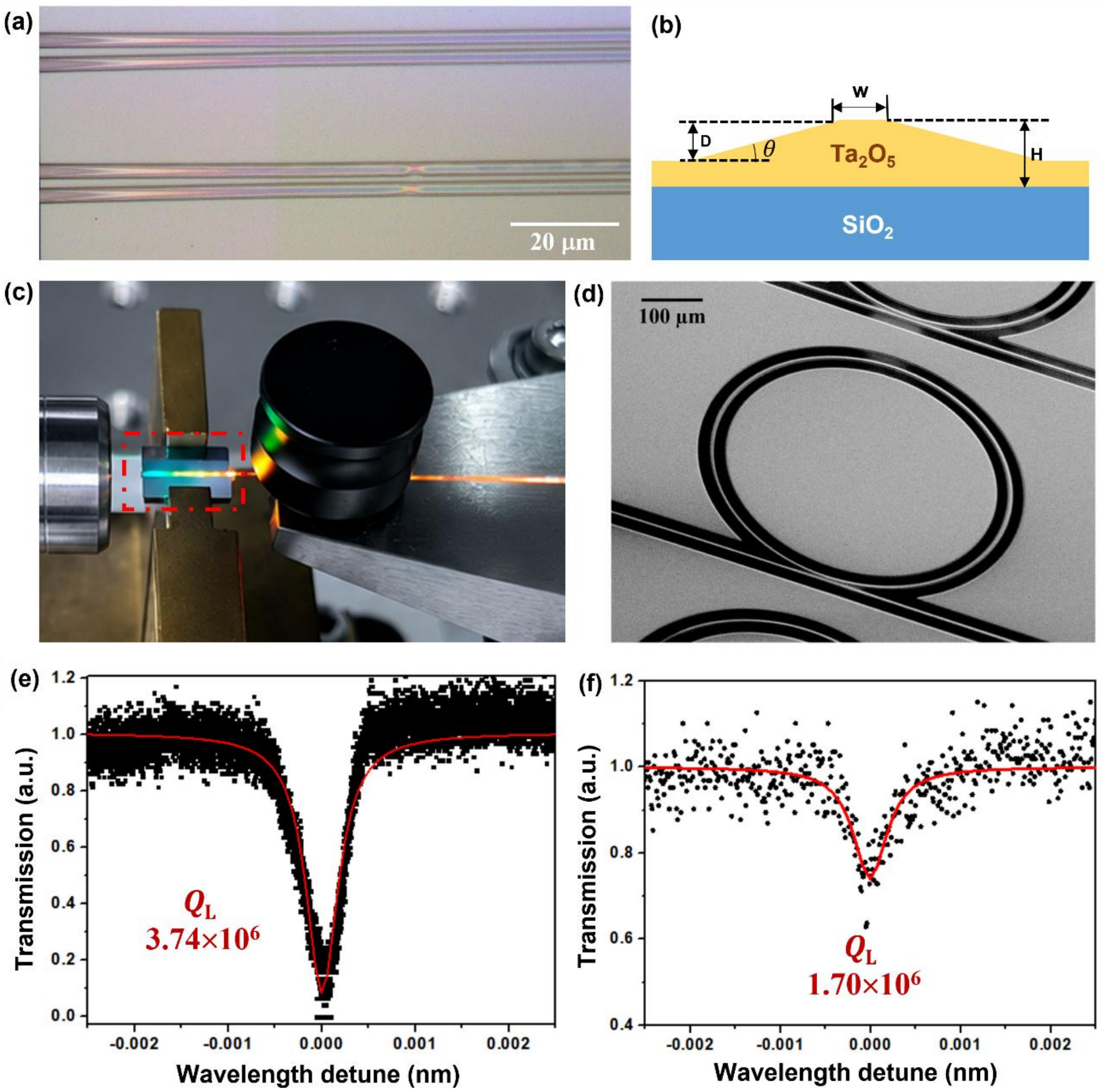

**Fig. 1. Nanophotonic waveguides and microring resonators.** (**a**) Optical micrograph of the fabricated waveguides near the end facet. (**b**) Schematic of the cross-section of the rib waveguide. (**c**) Photograph of the waveguide chip under test, with the chip highlighted by a red box. (**d**) Scanning-electron-microscope (SEM) image of the fabricated microring. (**e**) Loaded Q factor measured around 1551 nm. (**f**) Loaded Q factor measured around 780 nm.

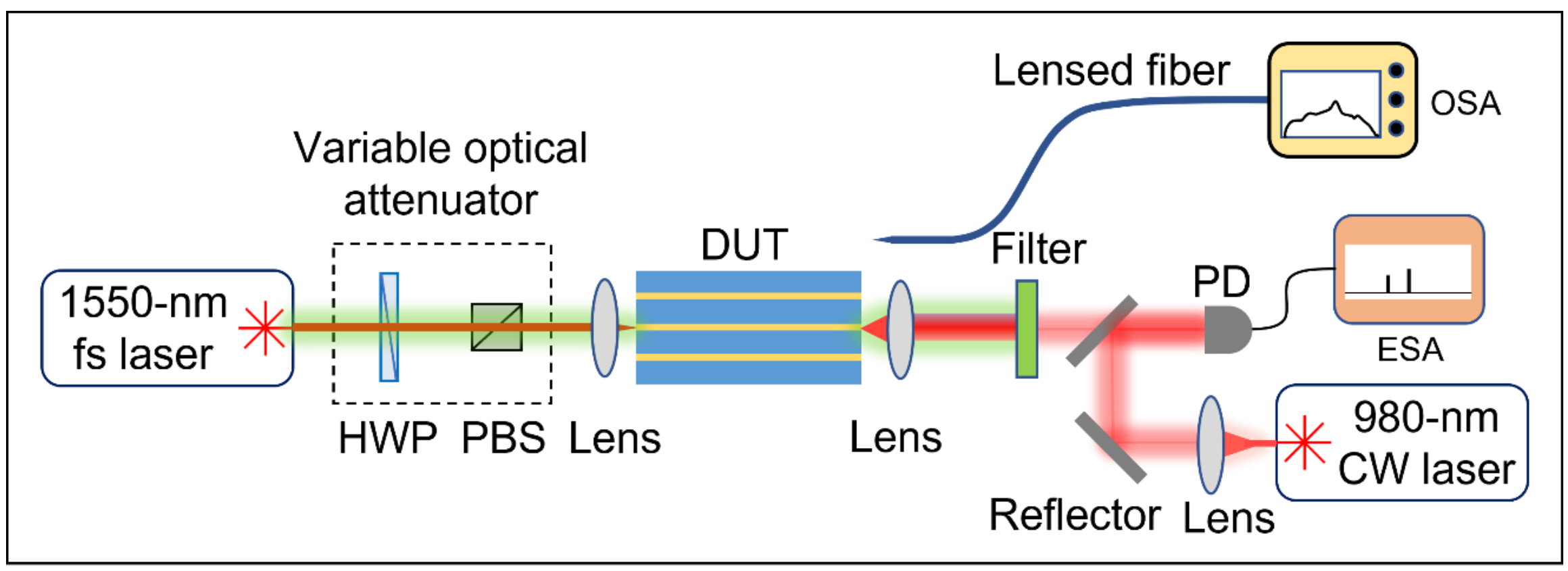

**Fig 2. Experimental setup for supercontinuum generation in the waveguides.** HWP: half-wave plate; PBS: polarizing beam-splitter; DUT: device under test; OSA: optical spectrum analyzer; PD: photodetector; ESA: electrical spectrum analyzer; fs: femtosecond: CW: continuous-wave.

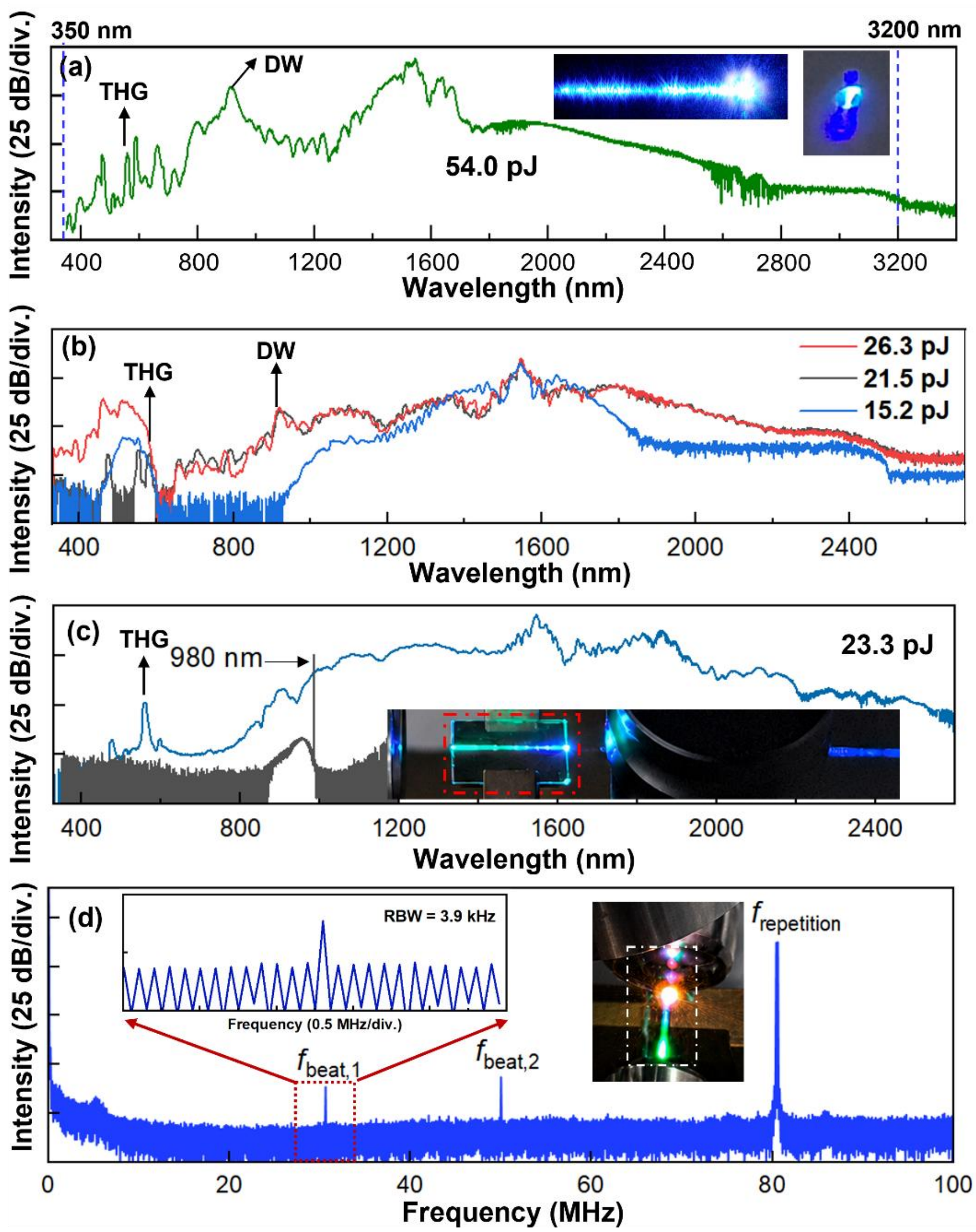

**Fig. 3. Supercontinuum generation in the waveguides.** (**a**) Optical spectrum of the ultra-broadband supercontinuum generated in the anomalous-dispersion waveguide. **Inset** (left): photograph of the scattered supercontinuum light taken over the entire waveguide. **Inset** (right): photograph of the output supercontinuum projected onto a sheet of blank white paper. (**b**) Spectral evolution of the supercontinuum with increasing pump pulse energy. (**c**) Optical spectrum of the flat supercontinuum in the normal-dispersion waveguide, together with the spectrum of the narrow-linewidth reference laser (black curve). **Inset**: photograph of the scattered supercontinuum emission along the waveguide. (**d**) Ratio-frequency spectrum of the heterodyned beat notes with a resolution bandwidth (RBW) of 3.9 kHz. **Inset** (left): zoom-in view of a single beat note. **Inset** (right): photograph of the waveguide emission taken from another viewing angle.

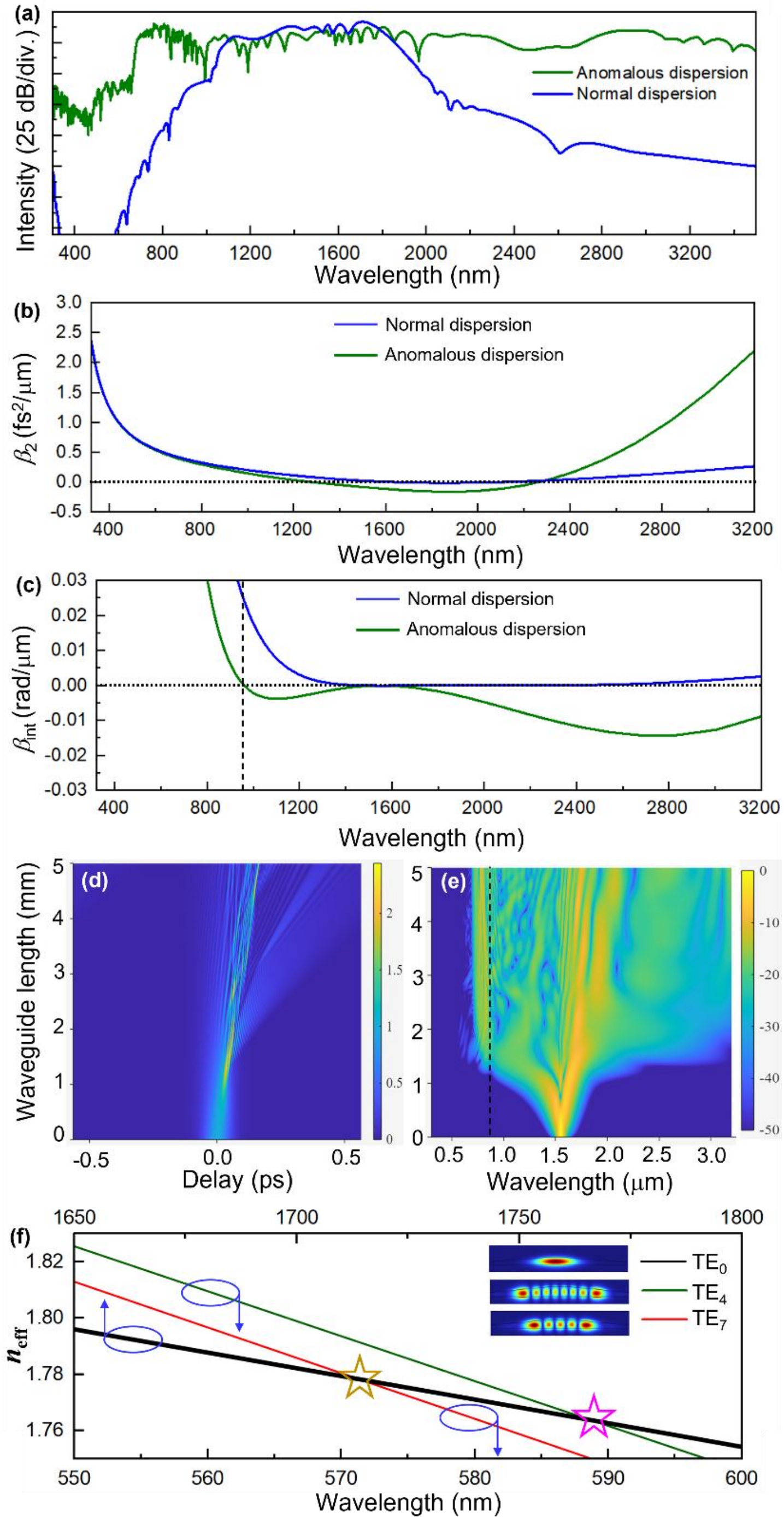

**Fig. 4. Dispersion, nonlinear dynamics, and phase-matching analysis.** (**a**) Simulated supercontinuum spectra for the anomalous-dispersion and normal-dispersion waveguides. (**b**) Group velocity-dispersion ($\beta_2$) and (**c**) integrated dispersion ($\beta_{int}$) of the two waveguides. (**d**) Temporal and (**e**) spatial evolution of the pump pulse along the propagation distance in the anomalous-dispersion waveguide. (**f**) Phase-matching analysis for third-harmonic generation (THG).

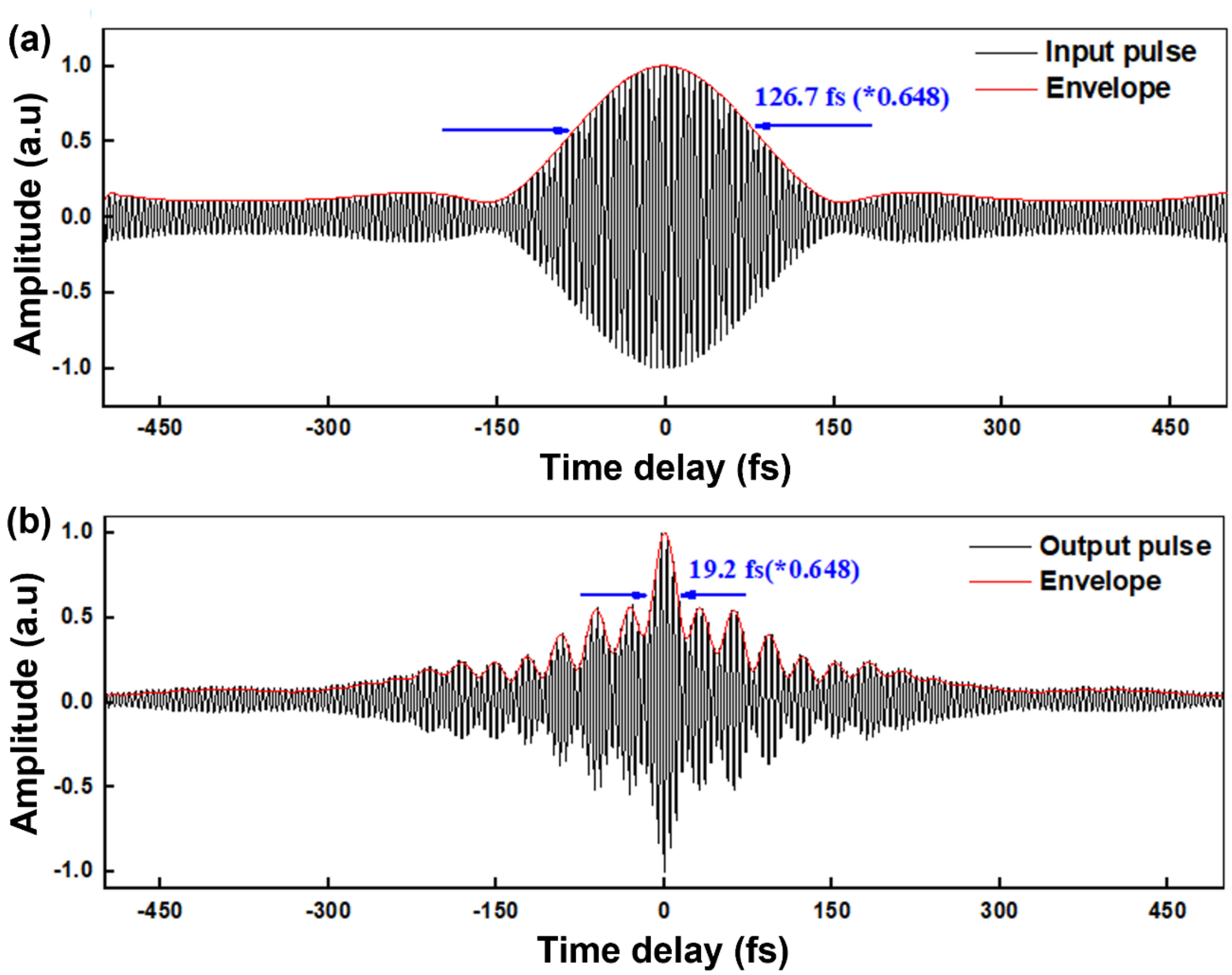

**Fig. 5. Soliton self-compression.** Field autocorrelation traces of (**a**) the incident pump pulse and (**b**) the compressed output pulse.